%% file: R_Article.tex
\documentclass[twocolumn,secnumarabic,amssymb, nobibnotes, aps, prb,superscriptaddress]{revtex4-1}

\usepackage{graphicx}
\usepackage{units}   
\usepackage{amsmath,nicefrac}
\usepackage{longtable}
\usepackage{float}
\usepackage[colorlinks=true, linkcolor=blue, citecolor=blue, filecolor=blue, urlcolor=blue]{hyperref}
\usepackage{changes}
\usepackage{cleveref}

\usepackage{epstopdf}

\begin{document}

\title{
	Wavelength-dependent reflectivity changes on gold at elevated electronic temperatures}

\author{A. Blumenstein}
\affiliation{ Laser-Laboratorium G\"ottingen e.V., Hans-Adolf-Krebs-Weg 1, D-37077 G\"ottingen, Germany}	
\affiliation{University of Kassel, Heinrich-Plett-Stra\ss{}e 40, D-34109 Kassel, Germany}	

\author{E.S. Zijlstra}	
\affiliation{University of Kassel, Heinrich-Plett-Stra\ss{}e 40, D-34109 Kassel, Germany}	

\author{D.S. Ivanov}	
\affiliation{University of Kassel, Heinrich-Plett-Stra\ss{}e 40, D-34109 Kassel, Germany}
\affiliation{Department of Physics and OPTIMAS Research Center, Technische Universität Kaiserslautern, Erwin-Schr\"odinger-Stra\ss{}e 46, D-67663 Kaiserslautern, Germany}

\author{S.T. Weber}
\affiliation{Department of Physics and OPTIMAS Research Center, Technische Universität Kaiserslautern, Erwin-Schr\"odinger-Stra\ss{}e 46, D-67663 Kaiserslautern, Germany}

\author{T. Zier}	
\affiliation{University of Kassel, Heinrich-Plett-Stra\ss{}e 40, D-34109 Kassel, Germany}

\author{F. Kleinwort}
\affiliation{ Laser-Laboratorium G\"ottingen e.V., Hans-Adolf-Krebs-Weg 1, D-37077 G\"ottingen, Germany}

\author{B. Rethfeld}	
\affiliation{Department of Physics and OPTIMAS Research Center, Technische Universität Kaiserslautern, Erwin-Schr\"odinger-Stra\ss{}e 46, D-67663 Kaiserslautern, Germany}

\author{J. Ihlemann}
\affiliation{ Laser-Laboratorium G\"ottingen e.V., Hans-Adolf-Krebs-Weg 1, D-37077 G\"ottingen, Germany}

\author{P. Simon}
\affiliation{ Laser-Laboratorium G\"ottingen e.V., Hans-Adolf-Krebs-Weg 1, D-37077 G\"ottingen, Germany}

\author{M.E. Garcia}	
\affiliation{University of Kassel, Heinrich-Plett-Stra\ss{}e 40, D-34109 Kassel, Germany}

\date{\today}%


\begin{abstract}
Upon the excitation by an ultrashort laser pulse the conditions in a material can drastically change, altering its optical properties and therefore the relative amount of absorbed energy, a quantity relevant for determining the damage threshold and for developing a detailed simulation of a structuring process. The subject of interest in this work is the d-band metal gold which has an absorption edge marking the transition of free valence electrons and an absorbing deep d-band with bound electrons. Reflectivity changes are observed in experiment over a broad spectral range at ablation conditions. To understand the involved processes the laser excitation is modeled by a combination of first principle calculations with a two-temperature model. The description is kept most general and applied to realistically simulate the transfer of the absorbed energy of a Gaussian laser pulse into the electronic system at every point in space at every instance of time. An electronic temperature-dependent reflectivity map is calculated, describing the out of equilibrium reflectivity during laser excitation for photon energies from $\unit[0.9-6.4]{eV}$, including inter- and intra-band transitions and a temperature-dependent damping factor. The main mechanisms are identified explaining the electronic temperature-dependent change in reflectivity: broadening of the edge of the occupied/unoccupied states around the chemical potential $\mu$, also leading to a shift of $\mu$ and an increase of the collision rate of free s/p-band electrons with bound d-band holes. 
\end{abstract}
\maketitle
\input{introduction}

\input{physical_picture}
\input{DFT}
\input{R_Article_Room_temperature}

\input{R_Article_elevated_temperatures}

\input{experiment}

\input{TTM}
\input{comparison}
\input{conclusion}

\begin{acknowledgements}
The present work was supported by the Deutsche Forschungsgemeinschaft (DFG) grants IH 17/18-1, IV 122/1-1, IV 122/1-2, RE 1141/14, RE 1141/15 and 600 GA 465/15-1, as well as, the Carl-Zeiss Foundation. We thank V. Roddatis from University of G\"ottingen for TEM measurements, determining the sample thickness. The calculations for this work were performed on Lichtenberg Super Computer Facility within the project 242.
\end{acknowledgements}

\bibliography{JabRef_Databank_Reflectivity}

\end{document}

%% file: introduction.tex
\section{Introduction}
The unique visible appearance of gold has fascinated not only scientists since its discovery. The origin of its typical colorful metallic look lies in the electronic properties resulting from the position of the absorbing d-band approximately \unit[2.35]{eV} below the Fermi level $E_F$, which leads to a high reflection of photons in the red to yellow spectral range and to a low reflection of photons in the blue to purple colored frequency interval. At photon energies too low to excite electrons from the d-band, the Drude-model can be applied \cite{drude1900elektronentheorie}. It describes the s/p-band electrons as freely oscillating in an electron gas resulting in a high reflectivity, typical for a metal and allows the use of gold as a reflective coating in the infrared (IR) energy range. Higher photon energies can excite bound electrons from the d-band leading to a strong increase in the absorption.
\\
The extreme conditions during strong laser excitation however can influence the electronic system in a material and thus change its reflection/absorption behavior. Especially at the non-equilibrium conditions necessary for a precise surface structuring of a metal where the laser energy is introduced below the electron-phonon relaxation time in a picosecond regime \cite{corkum1988thermal,mueller2013relaxation}. These extreme conditions lead to electrons at elevated temperatures in a relatively cold lattice and can be described by a two-temperature model (TTM) \cite{anisimov1974electron}. Pump-probe experiments around ablation conditions confirm that the electronic temperature $T_e$ is the main parameter determining transient reflectivity changes  \cite{hohlfeld2000electron,sun1994femtosecond,schoenlein1987femtosecond,fourment2014exp,ping2006broadband}. They show that the reflectivity, even after the fast heating of the electronic system by a pump pulse is staying in a transient state, due to the delayed transfer of energy to the lattice and its related decrease in electronic temperature. \\
A complete picture of the wavelength dependent reflectivity changes around the absorption edge of gold and near the ablation threshold is to our knowledge not presented yet. What has been done are thermo-reflectance studies describing the effect very close to the absorption edge and only up to $\unit[T_e=4]{kK}$ \cite{hohlfeld2000electron,sun1994femtosecond,schoenlein1987femtosecond}. Other studies describe the changes in $R(T_e)$, by describing in detail the effects valid for transitions within the s/p-band \cite{petrov2013thermal} up to $\unit[T_e=80]{kK}$ \cite{lin2008electron,holst2014ab} and in agreement with experiments \cite{fourment2014exp,zhang2015modeling} but only around the excitation energy $\unit[\hbar \omega=1.55]{eV}$. Also the conditions far above the ablation threshold of gold $\unit[F_{inc}=0.2-2]{Jcm^{-1}}$, \cite{preuss1995sub,furusawa1999ablation} with conditions $\unit[T_e > 100]{kK}$ are well studied allowing the use of a plasma state model, applicable for describing the reflectivity for a wide class of materials and wavelength \cite{milchberg1989light,fedosejevs1990absorption,price1995absorption,eidmann2000hydrodynamic}. However at the parameters usually used for metal surface structuring a plasma state is not reached and the model can not be applied. \\
The reflectivity map $R(T_e,\hbar \omega)$ of gold calculate in this work covers typical surface structuring conditions. It allows a precise description of the transient change of reflected and absorbed energy during the excitation by a laser pulse for a variety of parameters. With the limitations of the calculation of a maximum electronic temperatures of $\unit[T_e\approx80]{kK}$ and photon energies from $\unit[0.9-6.4]{eV}$ ($\unit[1380-190]{nm}$). Therefore the electronic system of gold is modeled by density functional theory (DFT) using the WIEN2k code \cite{ambrosch2006linear}. A modification of the code allows an ab-initio calculation of the density of states (DOS) under elevated electronic temperature conditions. At an increased $T_e$ also the effect of a change in the occupation of states is taken into account as well as a change of the electron-hole (eh)-collision rate entering the Drude-term \cite{fourment2014exp,zhang2015modeling,petrov2013thermal}. \\
A problem one faces now is to precisely relate the calculated reflectivity map $R(T_e,\hbar \omega)$ to experimental results since the transient state of $T_e$ in a bulk material is a parameter difficult to control experimentally and also to simulate in a model. A common way to overcome this problem is the use of thin gold foils \cite{ping2010warm,chen2013evolution} or films \cite{hohlfeld2000electron,sun1994femtosecond,schoenlein1987femtosecond} (\unit[10]{nm} to \unit[100]{nm}) in which the electronic temperature is homogeneously distributed in the probed area of the layer due to fast ballistic electrons. Another common approach is the use of a 1-D model simulating the electronic temperature only in depth assuming negligible difference in the incident laser fluence along the lateral direction \cite{fourment2014exp, fedosejevs1990absorption}.\\
For experiments with self-reflecting laser pulses on a bulk target and for determining the introduced energy for structuring processes \cite{ivanov2015experimental, wu2014microscopic}, a different approach is needed. Here{\color{red},} we provide a state of the art model being able to describe an arbitrary pulse with a symmetry around the $z$-Axis, reflected on a gold bulk target surface (thick film $\unit[>400]{nm}$) with a pulse length shorter than the characteristic electron-phonon equilibration time \cite{corkum1988thermal}. To describe the heat transport after absorption in detail, a TTM is used including the electron-phonon relaxation \cite{anisimov1974electron}. The symmetry around the $z$-axis is used to reduce the simulation to a 2-D model of the cross section of the pulse shape. In addition ballistic electrons are considered leading to an effective increase of the laser energy deposition depth. In the conducted experiments, pulse energies from $\unit[0.3]{\mu J}$ to $\unit[500]{\mu J}$ for $\unit[\hbar \omega=1.66]{eV}$ ($\unit[745]{nm}$) and $\unit[\hbar \omega=4.98]{eV}$ ($\unit[248]{nm}$) are applied.\\
The combination of this macroscopic description of energy transfer during laser pulse self-reflection by a TTM combined with the microscopic calculation of the excitation in the crystal based on DFT, gives us the possibility to directly relate the parameter of electronic temperature to the reflectivity, applicable to a wide variety of geometries and parameter sets. Allowing also a description of the physical phenomena leading to the reflectivity changes under electron phonon non-equilibrium conditions.

%% file: physical_picture.tex
\section{Physical picture of integrated laser pulse reflection}
In the following a general theoretical description of the integrated reflectivity is given, resulting from the transient process of a strong ultra-short laser pulse reflected by a bulk material surface. For a realistic simulation of the phenomenon the precise evolution of the electronic temperature during the self-reflection and the microscopic understanding of the electron dynamics under elevated $T_e$, at a certain photon energy $\hbar \omega$  in the bulk and its effect on the reflectivity are crucial. Experimentally the reflectivity is simply determined by the reflected laser pulse energy $E_\mathrm{ref}(T_e,\hbar \omega )$ divided by the incident laser pulse energy $E_\mathrm{inc}$. This gives the integral reflectivity $R_\mathrm{int}(T_e,\hbar \omega)$ of the pulse since every photon is reflected at a slightly different location and time and thus is related to a different electronic temperature $T_e$. For a theoretical model $R_\mathrm{int}(T_e,\hbar \omega)$ can also be expressed as a function of the absorbed energy $E_\mathrm{abs}(T_e,\hbar \omega)$    
\begin{eqnarray}
R_\mathrm{int}(T_e,\hbar \omega) = \frac{E_\mathrm{ref}(T_e,\hbar \omega)}{E_\mathrm{inc}} = 1-\frac{E_\mathrm{abs}(T_e,\hbar \omega)}{E_\mathrm{inc}} \enspace .
\label{eq:reflect}	
\end{eqnarray}
$E_\mathrm{abs}(T_e,\hbar \omega)$ can be obtained by integrating the source term $S(r,z,t,T_e,\hbar \omega)$ of the laser pulse: over the radius $r$, the depth $z$, the time $t$ and around the angle $\varphi$, assuming that the incident laser pulse has a symmetry around the $z$-Axis, 
\begin{eqnarray}
E_\mathrm{abs} (T_e,\hbar \omega) = \frac{1}{2 \pi} \int\limits_{0}^{\infty}\int\limits_{0}^{h}\int\limits_{0}^{\frac{d}{2}}\int\limits_{0}^{2\pi} S(r,z,t,T_e,\hbar \omega) d\varphi rdrdzdt \enspace ,
\label{eq:absorbed}	
\end{eqnarray}
with the simulation height $h$ and diameter $d$. The laser source describes the absorbed intensity distribution by the material at any point in space and time, as a function of the electronic temperature $T_e$ for a given $\hbar \omega$ at the surface and by the incident intensity distributions given by the functions: $f_r(r)$ along the radial distribution, $f_z(z)$ in depth and $f_t(t)$ along the temporal shape, 
\begin{eqnarray}
S(r,z,t,T_e,\hbar \omega) = E_\mathrm{inc}(1-R(T_e,\hbar \omega)) f_r(r) f_z(z,T_e,\hbar \omega) f_t(t) \enspace .
\label{eq:source}
\end{eqnarray}
The energy deposition by $S(r,z,t,T_e,\hbar \omega)$ into the electronic subsystem, its precise evolution over time and the coupling to the lattice can now be obtained by solving the differential equations of a TTM. Described in detail in section~\ref{sec:TTM}. With this precise description of the parameter $T_e$ depending on the incident laser pulse, the next step is to relate the reflectivity change to $T_e$. \\
Under normal conditions however at the surface between vacuum and a material the reflectivity is 
\begin{eqnarray}
R = \frac{(n_1-1)^2+k_1^2}{(n_1+1)^2+k_1^2} \enspace ,
\label{eq:reflectivity}	
\end{eqnarray} 
given by the refractive index of the material  
\begin{eqnarray}
n_1=\sqrt{\frac{|\epsilon(\omega)|+\epsilon^{\prime}(\omega)}{2}} ~\textnormal{and}~ k_1=\sqrt{\frac{|\epsilon(\omega)|-\epsilon^{\prime}(\omega)}{2}} \enspace ,
\label{eq:n_and_k}	
\end{eqnarray} 
the extinction coefficient, respectively. The complex dielectric function is given by $\epsilon (\omega) = \epsilon^{\prime}(\omega) + i \epsilon^{\prime \prime}(\omega)$. It describes the permittivity of a material for electro-magnetic waves, and is determined by the configuration of the valence electrons in the bulk. In gold excitations are described in a model of free- and bound electrons by inter-band transitions (between the d- and s/p-band) and intra-band transitions (within the s/p-band), respectively. The different excitations can be expressed separated by the dielectric function depending on the frequency $\omega$ of the incident photons \cite{ambrosch2006linear}, 
\begin{eqnarray}
\epsilon (T_e, \omega) = \epsilon^{\{\mathrm{inter}\}} (T_e, \omega) + \epsilon^{\{\mathrm{intra}\}} (T_e, \omega) \enspace .
\label{eq:wien2k}	
\end{eqnarray}
and is extended by the out of equilibrium description by $T_e$ relevant under the extreme conditions of laser excitation when the DOS and also its occupation of states can change having significant influence on the dielectric function. The inter-band part describing excitations from bound to unbound states can be obtained by calculating the imaginary part of the inter-band contribution to the dielectric tensor given by:
\begin{widetext} 
\begin{equation}
\epsilon_{ij}^{\{\mathrm{inter}\}}(T_e, \omega) = \frac{\hbar^2 e^2}{\pi m^{*2} \omega^2} \sum_{n,n'} \int \limits_{\textbf{k}} p_{i;n',n,\textbf{k}} ~ p_{j;n',n,\textbf{k}} \left( f_0(T_e,E_{n,\textbf{k}}) -  f_0(T_e,E_{n',\textbf{k}})\right) \delta(E_{n',\textbf{k}}-E_{n,\textbf{k}} -\omega)  \enspace .
	\label{eq:inter}	
\end{equation}
\end{widetext} 
it is determined by $\textbf{p}_{n',n,\textbf{k}}$ the momentum transition-matrix elements describing the excitation probabilities from a bound state in the band $n'$ to a free state in the band $n$ with a crystal momentum $\textbf{k}$. Also including the electronic temperature dependent Fermi-Dirac distribution $f_0(T_e)$ in which the single particle energies in the bands are described by $E_\textbf{k}$. Its derivation and precise calculation method by DFT is described for equilibrium conditions in detail by Ambrosch-Draxl et al. \cite{ambrosch2006linear}.\\
The intra-band part $\epsilon^{\{\mathrm{intra}\}}(T_e, \omega)$, describing the transitions within the free electron gas of the metal, can be described in the Drude-model \cite{drude1900elektronentheorie,johnson1972optical}
\begin{eqnarray}
\epsilon^{\{\mathrm{intra}\}} (T_e, \omega) = 1 - \frac{\omega_{p;ij}^2}{\omega^2 + i \nu_e (T_e) \omega} \enspace .
\label{eq:drude}	
\end{eqnarray}
Using the WIEN2k code, a plasma frequency 
\begin{equation}
\omega_{p;ij}^2 = \frac{\hbar^2 e^2}{\pi m^{*2} } \sum_{n} \int \limits_{\textbf{k}} p_{i;n,n,\textbf{k}} ~ p_{j;n,n,\textbf{k}}  \delta(E_{n,\textbf{k}}-E_F)  \enspace ,
\label{eq:plasma}	
\end{equation}
can be extracted as described by Ambrosch-Draxl et al. \cite{ambrosch2006linear}. In the Drude-Model the response of the free electrons in the model on the external laser field with $\omega$ is additionally damped  by the collision rate $\nu_e (T_e)$ describing collisions of the free electrons with bound electron states closer to the core. 

%% file: DFT.tex
\section{DFT Calculation of Broadband Reflectivity Change at Elevated Electronic Temperature}
The dielectric function $\epsilon (T_e, \omega)$ of crystalline gold is obtained by first principles DFT calculations using WIEN2k (Version 13.1) \cite{ambrosch2006linear}. The code is modified and extended to allow the calculation of an electronic temperature-dependent reflectivity value under electron-phonon non-equilibrium conditions. In WIEN2k, the method of all-electron full-potential linearized augmented plane waves is used to calculate the Kohn-Sham eigenstates, taking into account the screening effect of inner electrons, the influence of relativistic electrons close to the core, as well as the effect of spin-orbit coupling for optical transitions \cite{glantschnig2010relativistic}. The calculations of gold were performed in the local density approximation where we used $17$ valence orbitals and a lattice parameter of $0.408$ nm. The basis of $734$ plane waves to describe the valence electrons was determined by a maximal value
$k_{max}$ by $R*k_{max} = 8.0$, where $R$ is the radius of the used muffin tins. In
total the electronic band structure was calculated at $816$ $k$-points. To account for the width of the discrete energy levels a lifetime broadening of $\unit[\Gamma=10]{fs}$ is used.
\begin{figure}[h]
	\centering
	\includegraphics[width=0.46\textwidth]{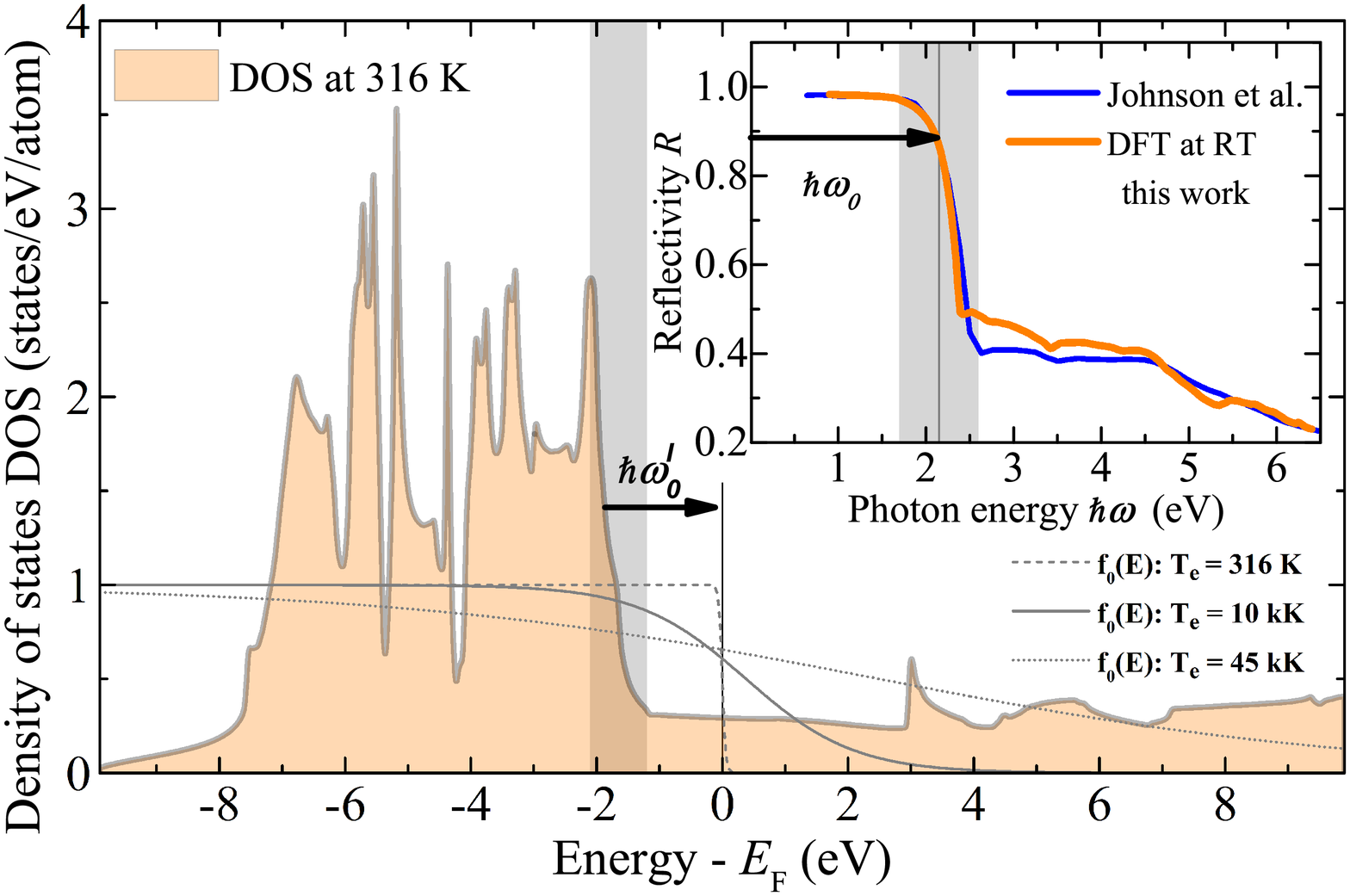}
	\caption{DOS of the valence electrons of crystalline gold with the atomic shell configuration [Xe]4f$^{14}$5d$^{10}$6s$^1$ calculated at RT with WIEN2k. The broad d-band edge of $\unit[0.9]{eV}$ width is highlighted in gray. The Fermi-Dirac distribution for different $T_e$ is included, and shifted by $ \Delta \mu (T_e)$. The obtained RT reflectivity is shown in the inset compared to the literature value from Johnson et al. \cite{johnson1972optical}.}
	\label{fig:DOS_R_Au_sim}
\end{figure}

%% file: R_Article_Room_temperature.tex
\subsection{Room Temperature conditions}

The WIEN2k code determines the DOS by ab-initio calculation and from that the inter-band transitions (Eq. (\ref{eq:inter})), the influence of these transitions are normally described in a Drude-model as the core polarization term \cite{fourment2014exp}. The intra-band transitions however given in the model from Ambrosch-Draxl et al. also use a Drude-term which requires a damping parameter at nearly room temperature (RT) here defined as \unit[316]{K}(\unit[0.002]{Ry}). In addition DFT calculation can not precisely describe the correct absolute energy positions of the Fermi level. Therefore the literature data for the dielectric function and thus $R(\hbar \omega)$ from Johnson et al. \cite{johnson1972optical} are used to define these two parameters at RT conditions. To obtain the correct absolute energy positions the calculated DOS at RT, shown in Fig.~\ref{fig:DOS_R_Au_sim} needs to be shifted by $\unit[\Delta E = 0.41]{eV}$. The reflection edge is defined at the DOS spanning over $\unit[0.9]{eV}$ marked in gray from the onset of the d-band to the first peak in its DOS. In the inset showing the shifted reflectivity it spans from $\unit[1.7]{eV}$ to $\unit[2.6]{eV}$ shaded also in gray in Fig.~\ref{fig:DOS_R_Au_sim}. At the center of this gray bar an photon energy of $\unit[\hbar \omega_0 \approx 2.15]{eV}$ is needed to excite in an unoccupied state above $E_F$ and marks the transition from high reflection ($\unit[R\approx0.97]{}$) for excitations from the s/p-band to low reflection ($\unit[R\approx0.34]{}$) for photon excitation from the d-band. The second parameter, the Drude-damping parameter $\unit[\nu_e(T_{e;RT}) = \nu_{eph}]{}$. which describes the electron phonon collision rate at RT conditions is defined by the reflectivity value given by literature at an photon energy of $\unit[1.66]{eV}$. Defining in our model the Drude-damping parameter to a value of $\unit[\nu_{eph}=0.88]{fs^{-1}}$ where the inter-band part is given by Eq. (\ref{eq:inter}), and the plasma frequency is given by Eq. (\ref{eq:plasma}).

%% file: R_Article_elevated_temperatures.tex
\subsection{Elevated Electronic Temperatures}
\label{sec:DFT}
\begin{figure}
	\centering
	\includegraphics[width=0.49\textwidth]{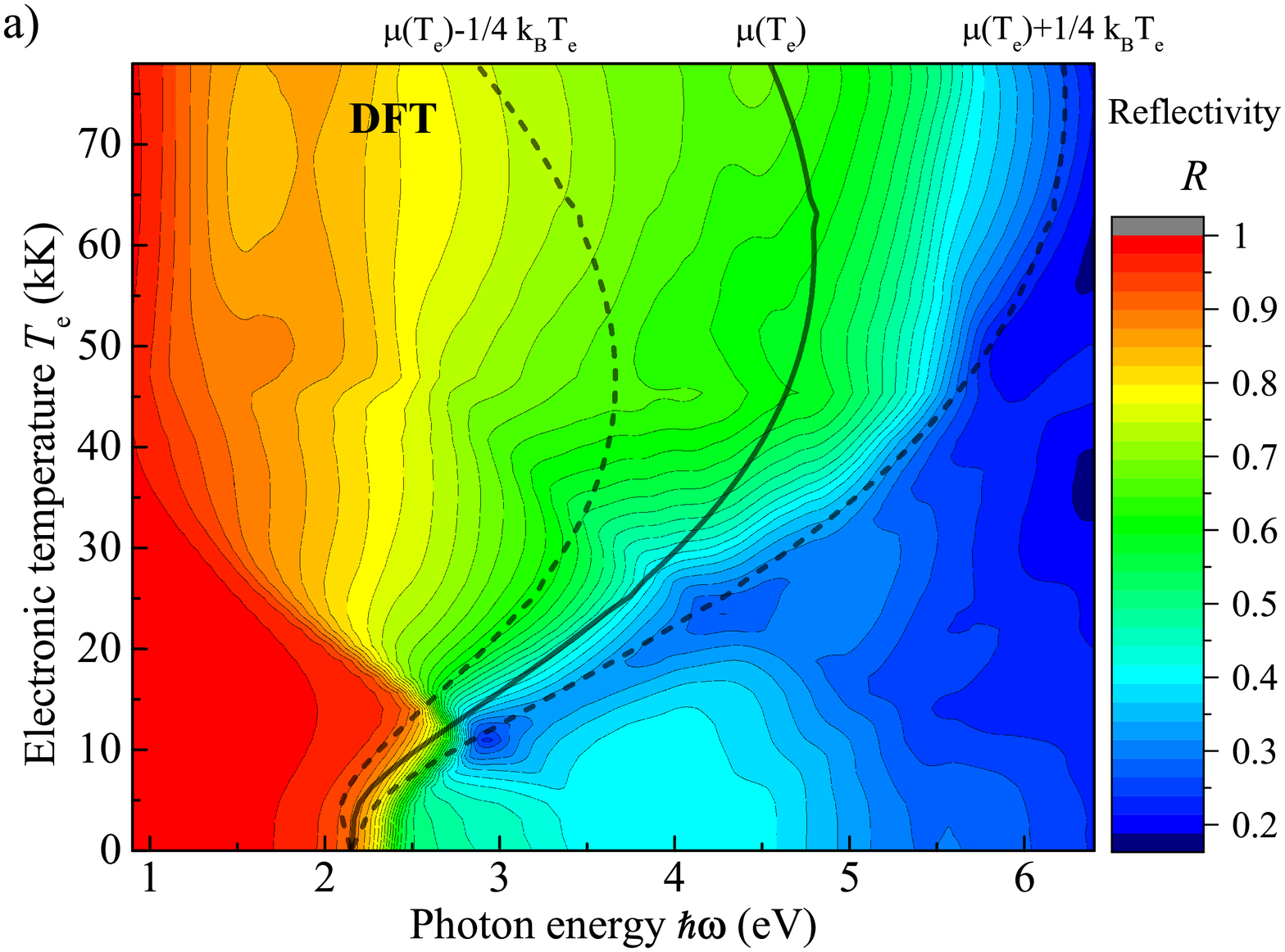}
	\includegraphics[width=0.49\textwidth]{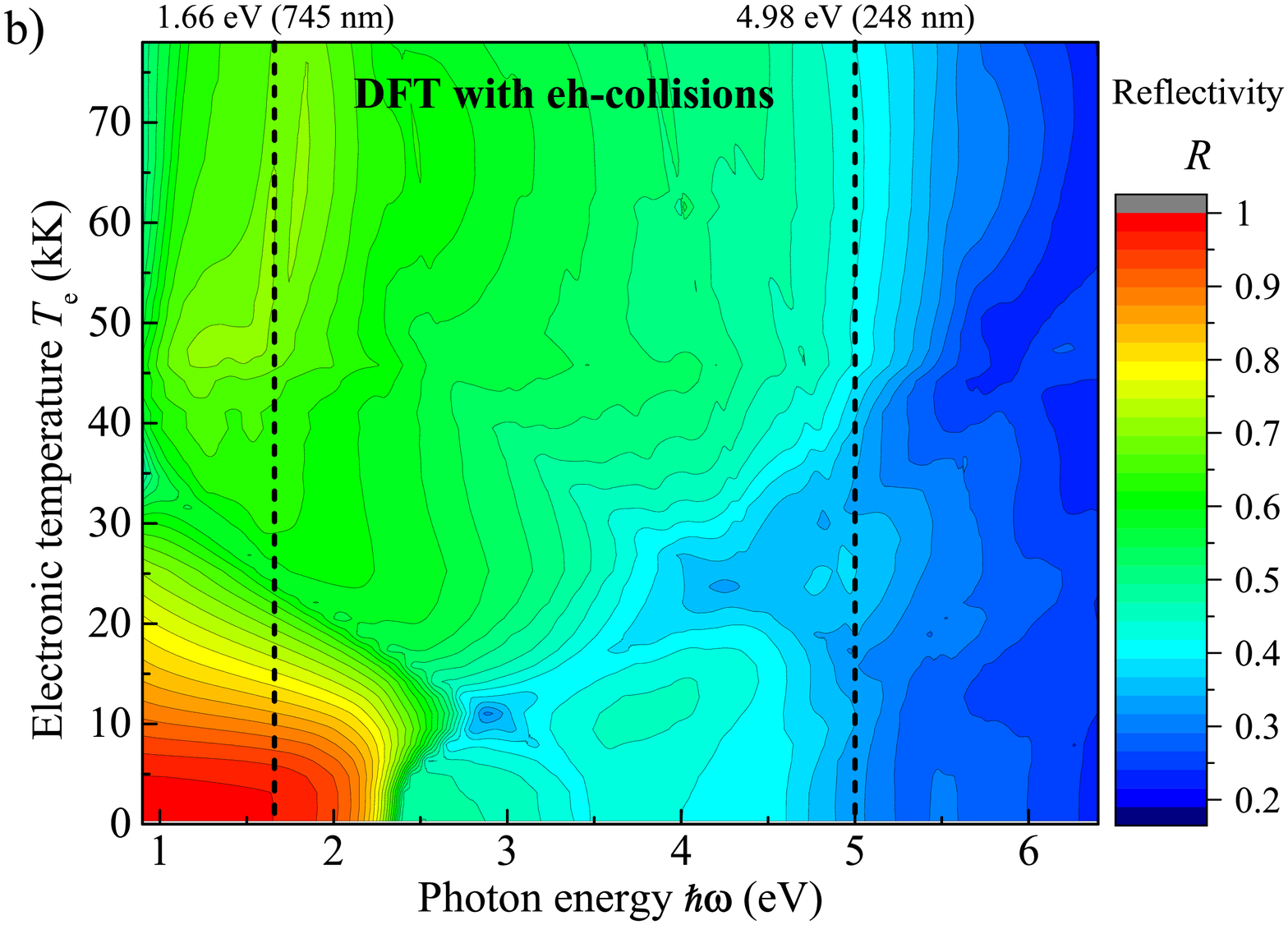}
	\caption{a) Calculated reflectivity map $R(T_e, \hbar \omega)$ in dependence of the electron temperature $T_e$ and photon energy $\hbar \omega$ obtained by DFT calculations using a temperature-dependent modification of the WIEN2k code. In b) the DFT calculations are extended by the effect of eh-collisions. Photon energies at \unit[1.66]{eV} and \unit[4.98]{eV} are marked with dotted lines at which the calculated results are compared to experiment.}
	\label{fig:R_map}
\end{figure}

The effect of an increase in the electronic temperature on the dielectric function introduced in Eqs.~(\ref{eq:wien2k}) and (\ref{eq:drude}) is implemented in our DFT simulation. First only the influence of the inter-band part and the plasma frequency shown in Fig. \ref{fig:R_map} a) are discussed. In a second step our model is extended by a temperature-dependent collision frequency $\nu_e(T_e)$ plotted in Fig. \ref{fig:R_map} b). 

In Fig. \ref{fig:R_map} a) (DFT) when considering a constant collision frequency of $\nu_e(T_{e;RT}) = \nu_{eph} = \unit[0.088]{fs^{-1}}$ over the whole reflectivity map shown in Fig. \ref{fig:R_map} (a) for the case of the electronic temperature near the equilibrium conditions the color plot changes from red to green in a range of $\unit[\approx 0.9]{eV}$. For increasing temperatures, the midpoint of the reflectivity edge however shifts to larger photon energies, also the chemical potential (thick gray line) shifts from $\unit[\hbar \omega_0 =2.15]{eV}$ up to $\unit[\hbar \omega_0 \approx 5]{eV}$ an effect described previously by Holst et al. and others \cite{holst2014ab}. It can be understood in the DOS picture where $\mu(T_e)$ is located so that the number of holes below and electrons above it are equal.
Since the DOS is higher below $\mu(T_e)$ a redistribution of electrons and holes due to a rising $T_e$ shifts the chemical potential towards higher energies. Another visible feature in Fig. \ref{fig:R_map} (a) is a broadening or smearing of the reflectivity edge mentioned previously by Ping at al. \cite{ping2010warm,ping2006broadband} from a range of $\unit[\approx 0.9]{eV}$ above electron temperatures to a broad edge ranging roughly from \unit[1.2]{eV} to \unit[6]{eV} at $\unit[T_e \approx 70]{kK}$. A possible explanation for this effect is a smearing introduced by the Fermi-Dirac distribution appearing only above electronic temperatures where the d-band is starting to depopulated, while the s/p-band is further populated. The onset temperature of this effect of \unit[10]{kK} matches roughly the width of the d-band edge itself shown in the DOS in Fig.~\ref{fig:DOS_R_Au_sim} (marked there in gray) and leads to an onset of the smearing only above ($\unit[T_e\approx10]{kK}$) $\unit[T_e\approx1]{eV}$.

In Fig. \ref{fig:R_map} b) (DFT with eh-collisions) in addition the effect of electron-hole-collisions on the reflectivity map $R(T_e, \hbar \omega)$ is included by introducing a $T_e$ dependent damping parameter \cite{petrov2013thermal,fourment2014exp,zhang2015modeling}. The results are compared with each other and later to experimental data. The dependent damping parameter enters Eq.\,(\ref{eq:drude}) as the collision frequency $\nu_e(T_e) = \nu_{eph} + \nu_{eh}(T_e) + \nu_{ee}$ which includes the electron-phonon collisions $\nu_{eph}$, representing the collisions also present at equilibrium conditions, extended by the dynamic part of the electron-hole collisions $\nu_{eh}(T_e)$ plus the effect of electron-electron collisions $\nu_{ee}$. The influence by electron-electron collisions $\nu_{ee}$ on the Drude damping term can be neglected when describing the dielectric function \cite{fourment2014exp}. 
The electron-hole-collision $\nu_{eh}(T_e)$ can be further divided in collisions with free holes in the s/p-band which play a minor role even at elevated electronic temperatures \cite{fourment2014exp} and the effect of collisions with bound holes in the d-band which do change the damping term in the Drude-model when $T_e$ is rising. The reason is the formation of unoccupied states in the d-band and an equal increase of occupied states in the s/p-band due to the Fermi-Dirac distribution and can be described by $N^{d}_h(T_e) = N_{e}^{sp}(T_e)-1$. These effective numbers of holes and electrons per atom in the d-band and sp-band respectively are calculated by the DOS from the DFT calculation. The collision frequency $\nu_{eh}(T_e)$ 
\begin{eqnarray}
\nu_{eh}(T_e) = A_{eh} N^{d}_e(T_e)N^{d}_h(T_e) \enspace ,
\label{eq:eh-collision}	
\end{eqnarray} 
is then given by the parameter $A_{eh}$ (assumed constant over $T_e$ and $\hbar \omega$) times the scattering electrons (which are the effective number per atom of all occupied states in the band 5d$^{10}$) times the holes in the d-band\cite{fourment2014exp}.

The two different reflectivity maps $R(T_e, \hbar \omega)$ can now be used to describe the effects on the transition edge, marking the change from high to low reflectivity values, at elevated electronic temperatures where we can separate the effect of the the ab-initio DFT calculations presented in Fig.~\ref{fig:R_map} a), from the effect of the additional inclusion of an $T_e$ dependent increase of eh-collisions shown in Fig.~\ref{fig:R_map} b). When comparing Fig.~\ref{fig:R_map} a) and b) the effect of the added dynamic damping term $\nu_{eh}(T_e)$ is visible especially in the IR, where the increase of $T_e$ leads to an earlier and stronger drop compared to the case without a temperature dependent damping term while the effect on the higher energetic photons is only a slight increase of the reflectivity. The results in Fig.~\ref{fig:R_map} b) below $\unit[T_e\approx10]{kK}$ around the absorption edge qualitatively agree with experimental thermo-reflectance results from the literature \cite{hohlfeld2000electron,schoenlein1987femtosecond,sun1994femtosecond}. Where a decrease in reflectivity below an excitation energy of $\unit[\hbar \omega_0 =2.35]{eV}$ and an increase above that excitation energy is measured, and related to a broadening of the edge. As an explanation for this broadening our calculation results suggest, that the dynamic damping term $\nu_{eh}(T_e)$ describing eh-collisions with the d-band explains this broadening, rather then the broadening of the excitation edge itself by a smearing of the edge of the occupied states in the s/p-band when excitation in a sharp d-band are assumed \cite{hohlfeld2000electron,schoenlein1987femtosecond,sun1994femtosecond}. In our ab-initio simulation (Fig.~\ref{fig:R_map} a)) we can see this effect, but only appearing above $\unit[T_e\approx10]{kK}$ and attribute this to the width of the d-band edge itself as described above. 
With this detailed knowledge of the effect of $T_e$ on the reflectivity map the next task is to relate this parameter with a model to the experiment. 

The maps from $R(T_e, \hbar \omega)$ with eh-collisions includes and also $\epsilon'(T_e, \hbar \omega)$, $\epsilon''(T_e, \hbar \omega)$, respectively are included in the supplementary material presented in matrix form.

%% file: experiment.tex
\section{Experimentally Determined Reflectivity}
\label{sec:exp}
The integral-reflectivity is obtained under common experimental conditions of a spatially and temporally shaped Gaussian laser pulse. The incident $E_\mathrm{inc}$ and reflected energy $E_\mathrm{ref}(T_e)$ is measured to obtain the integral reflectivity $R_\mathrm{int}(T_e)$ by the use of Eq.\,(\ref{eq:reflect}). Two different photon energies are probed \unit[1.66]{eV} (IR) and \unit[4.98]{eV} (UV) with a pulse length of $\unit[\tau^{IR}=0.6]{ps}$ and $\unit[\tau^{UV}=1.6]{ps}$ respectively. The pulse energy is varied between $\unit[0.3 ]{\mu J}$ and $\unit[500]{\mu J}$, to obtain typical structuring conditions covering the ablation threshold of gold in the UV $\unit[F_{inc}^{UV}\approx0.2]{Jcm^{-1}}$, \cite{preuss1995sub} and the IR $\unit[F_{inc}^{IR}\approx1.5]{Jcm^{-1}}$, \cite{furusawa1999ablation} which both show also a pulse length dependency, described in the present work by the parameter $T_e$. The IR experiments are conducted with pulses from a Ti:Sapphire regenerative amplifier system. For the UV experiments the Ti:Sapphire pulses were frequency tripled and amplified in a KrF excimer module to reach sufficient pulse energy \cite{bekesi2002table,szatmari1991off}. The UV pulses pass a \unit[3]{mm} aperture and are focused with a $\unit[f=1000]{mm}$ fused silica lens to a nearly Gaussian spot of $\unit[b=85]{\mu m}$ FWHM. For the IR experiments a $\unit[f=500]{mm}$ lens is used to create the same spot size for a larger Gaussian beam profile. An ultrashort single laser pulse is reflected under nearly normal incidence $<5^\circ$ by a thick gold target and the incident and reflected energy of the pulse are measured. A x-y-z-stage is used to move the sample in the beam waist and to allow every pulse to hit a new undamaged area of the gold sample. The polycrystalline gold films are deposited by evaporation technique on a polished glass surface with a \unit[50]{nm} chromium layer for enhanced lattice matching between gold and the substrate. The gold film has a thickness of \unit[400]{nm} and an estimated grain size of \unit[~75]{nm}, determined by analyzing the cross section of TEM measurements. To ensure a high precision of the data, four different energy detectors in overlapping scales are used: A Polytec RjP-735, Ophir PE10BF-C, Ophir PE9-ES-C and Ophir PD10-C. In the setup one pulse energy sensor measures the Fresnel-reflection of the focusing lens while a second sensor measures the reflected pulse and is calibrated prior to the measurement. The pulse length is determined using a home made single-shot UV-FROG described in \cite{nagy2009single}, and in the IR by a "Positive Light" single shot autocorrelator.

%% file: TTM.tex
\section{TTM Simulation of Integral Reflectivity}
\label{sec:TTM}
\begin{figure}[h]
  	\centering
  	\includegraphics[scale=0.27]{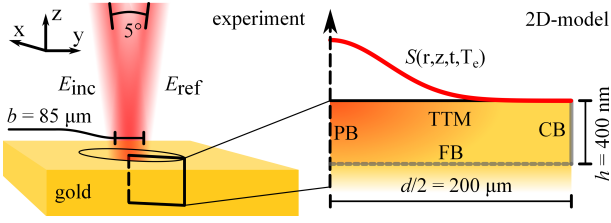}
  	\caption{Schematic picture of the nearly Gaussian laser focus spot with its experimental parameters. The inset shows the 2D-model geometry in cylindrical coordinates with the source term $S(r,z,t,T_e)$. The model boundary conditions are: At the $z$-axis a periodic boundary (PB), in depth free boundary (FB) and at the surrounding circle a constant boundary (CB).}
  	\label{fig:setup_model}
  \end{figure}

The connection between the reflectivity map $R(T_e,\hbar \omega)$ obtained in the DFT calculations and the experimentally determined integral value of self-reflectivity is realized by solving the time-dependent differential equations of a TTM simulation \cite{anisimov1974electron}. Therefore the temperature-dependent reflectivity is implemented in the source term as a tabulated value obtained from line profiles at the experimental probe photon energy from the map $R(T_e,\hbar \omega)$ in Fig \ref{fig:R_map} a) and b). The volume of interaction is modeled by a disk with a thickness of $\unit[h=400]{nm}$ and a diameter of $\unit[d=400]{\mu m}$ shown in Fig.~\ref{fig:setup_model}. The lateral size of the model was chosen so that the electronic temperature would not raise for more than $\unit[0.1]{\%}$ by the end of the pulse.
A Gaussian distribution of the pulse in the temporal and spatial domain is used, with the pulse length $\tau$ and the focus diameter $\unit[b]{}$ at FWHM with using the values of the pulse from experiment described in sec. \ref{sec:exp}.
\begin{eqnarray}
f_r(r)= \frac{1}{\pi b^2} \exp\left(-\frac{r^2}{b^2}\right) \enspace .
\label{eq:fr}
\end{eqnarray}
The effective energy attenuation function in depth $z$ for the pulses is given by: 
\begin{eqnarray}
f_z(z)= \frac{\exp\left(-\frac{z}{\lambda (T_e, \hbar \omega ) + \lambda_\mathrm{b}}\right)}{\left(1-\exp\left(-\frac{h}{\lambda (T_e, \hbar \omega )+\lambda_\mathrm{b}}\right)\right) \lambda (T_e, \hbar \omega ) + \lambda_\mathrm{b}}  \enspace .
\label{eq:fz}
\end{eqnarray}
With the characteristic penetration depth depending on the electronic temperature $T_e$ and the photon energy $\hbar \omega$
\begin{eqnarray}
\lambda (T_e, \hbar \omega ) = \frac{n_1 c}{\omega \epsilon''(T_e,\omega)} \enspace .
\label{eq:lambda_T_e}
\end{eqnarray}
Where the refractive index $n_1$ from Eq. \ref{eq:n_and_k} and the imaginary part of the dielectric function $\epsilon''(T_e,\omega)$ from  Eq. \ref{eq:inter} is inserted. The values are calculated by the described model including the effect of eh-collisions and inserted as tabulated data in the TTM. The values for the penetration depth at room temperature are for the probed wavelength  $\unit[\lambda_{RT}^{IR}=13.1]{nm}$ and $\unit[\lambda_{RT}^{UV}=13.0]{nm}$ while for the highest calculated electronic temperature of $\unit[T_e \approx 79]{kK}$ the values are $\unit[\lambda_{79 kK}^{IR}=15.5]{nm}$ and $\unit[\lambda_{79 kK}^{UV}=11.4]{nm}$ varying in between  not more than $\unit[9]{nm}$. \\
The Eq. \ref{eq:fz} includes the sample thickness $h$ which is needed to account for possible losses of a thin sample relevant for low pulse energies with significant range of ballistic electrons $\lambda_\mathrm{b}$. It describes the non-thermalized electrons quickly proceeding deeper into the material and their thermalization process by means of electron-electron-collisions which occurs at a greater distance than the optical penetration depth $\lambda (T_e, \hbar \omega )$. The ballistic transport therefore can effectively increase the laser energy deposition by one order of magnitude in depth described by $\lambda_\mathrm{b}$. Neglecting the kinetics of this process, we include the ballistic range into the source term, Eq.\,(\ref{eq:source}), as it was suggested by Hohlfeld et al. \cite{hohlfeld2000electron}.\\
The temporal shape of the pulse is given by:
\begin{eqnarray}
f_t(t)= \frac{1}{\tau} \sqrt{\frac{\sigma}{\pi}} \exp\left(-\sigma\frac{(t-t_0)^2}{\tau^2}\right) \enspace ,
\label{eq:ft}
\end{eqnarray}
with $\sigma = 4 \ln2$. The laser pulse is defined by its duration $\tau$. At $t = 0$ a single pulse shifted to $t_0  =  2.5 \tau$ is used to inscribe a Gaussian temporal profile with a length of \unit[1.6]{ps} and \unit[0.6]{ps} respectively. 

The inclusion of the ballistic transport can alter the electron temperature distribution in the target upon the laser pulse absorption. We estimate its additional temperature dependence according to the transport relaxation time via collision rates, $\tau_{rel}^{-1} = \tau_{ee}^{-1} + \tau_{eh}^{-1} + \tau_{eph}^{-1}$ \cite{landau1957theory,inogamov2009two}. The resulting relaxation time $\tau_{rel}$ will thus be a function of the electron, $T_e$, and phonon, $T_{ph}$, temperatures respectively with the corresponding contributions as $\tau_{ee}^{-1}\sim AT_e^2$ and $\tau_{eph}^{-1} \sim BT_{ph}$ \cite{anisimov1997theory,wang1994time}. The influence of $\tau_{eh}^{-1}$ can be neglected since the electron hole collisions take place only at temperatures when holes are created in the d-band at which the penetration depth is already below a few nm. With the value of ($\unit[1.39 \times 10^6]{m s^{-1}}$) for the Fermi-velocity of free electrons, in our simulations therefore the ballistic range $\lambda_\mathrm{b}$ in Eq.\,(\ref{eq:fz}) is dynamically changing during the pulse and ranges from \unit[300]{nm} at nearly zero intensity down to \unit[5]{nm} at the peak intensity, where the temperature of electrons reaches its maximum level. During our simulations, however, we noticed that the inclusion of the ballistic transport does not alter the integral reflectivity value for more than $\unit[0.1]{\%}$ in the range of incident energies greater than $\unit[20]{\mu J}$, which is in agreement with our previous publications \cite{hohlfeld2000electron,ivanov2015experimental} and theoretical predictions by Petrov et al. \cite{petrov2015two}. At lower incident energies, the ballistic range is comparable with the thickness of the modeled target and the first multiplier in Eq.\,(\ref{eq:fz}) is accounting for its finite size. This parametric description of the source $S(r,z,t,T_e)$ is now implemented in the TTM which in cylindrical coordinates can be written as:
\begin{widetext} 	
\begin{eqnarray}
    \left\{
	\begin{aligned}
	C_e(T_e) \frac{\partial T_e}{\partial t} &= \frac{1}{r}\frac{\partial}{\partial r}rK_e(T_e,T_a)\frac{\partial T_e}{\partial r}+  \frac{\partial}{\partial z}K_e(T_e,T_a)\frac{\partial T_e}{\partial z}- G(T_e)\left[T_e-T_a \right]+ S(r,z,t,T_e)  \label{eq:TTM_e} \\
	C_a \frac{\partial T_a}{\partial t} &= \frac{1}{r}\frac{\partial}{\partial r}rK_a\frac{\partial T_a}{\partial r}+ \frac{\partial}{\partial z}K_a\frac{\partial T_a}{\partial z}+ G(T_e)\left[T_e-T_a \right] \enspace ,
	\end{aligned}
 \right.
 \label{eq:TTM_a}
\end{eqnarray}
\end{widetext}
where indexes $e$ and $a$ are standing for the electrons and lattice correspondingly. Since working with a short laser pulse, the thermal transport due to phonon conductivity $K_a$ is negligible when compared to electron conductivity $K_e$ and thus the corresponding parts in Eq.\,(\ref{eq:TTM_a}) can be omitted. $G$ and $C$ are the strength of the electron-phonon coupling with the value for the lattice taken as $\unit[C_a=2.327]{MJK^{-1} m^{-3}}$ given by experimental data \cite{lide19921993}. By utilizing DFT calculated density of states with the effect of the d-band included, Lin et al. \cite{lin2008electron} determined the electron temperature dependence for the electron-phonon coupling and the electron heat capacity functions $G$ and $C_e$. These quantities therefore were considered in the present work in the form of tabulated data \cite{lin2008electron}. The complex behavior of the electron heat conductivity $K_e$ was approximated as it is suggested by Anisimov and Rethfeld \cite{anisimov1997theory,Rethfeld2017}, as a function of the electron and lattice temperatures $T_e$ and $T_a$ respectively:
\begin{eqnarray}
	K_e = \gamma \frac{(\theta^2_e + 0.16)^{5/4}(\theta^2_e + 0.44)\theta_e}{(\theta^2_e + 0.44)^{1/2}(\theta^2_e + \delta \theta_a)} \enspace ,
 \label{eq:K_e}
\end{eqnarray}
with the parameters $\theta_e=k_B T_e E_F^{-1}$, $\theta_a=k_B T_a E_F^{-1}$, $\unit[\gamma = 353]{W m^{-1}K^{-1}}$ and $\unit[\delta=0.16]{}$ for gold. This dependence shows linear behavior with $T_e$ at low excitation level, a significant decay at the excitation level comparable with $T_F$, and a steep increase like plasma conductivity at a higher excitation level. \\
Finally, during the simulations the laser pulse, centered at $r = 0$, is irradiating the target At the front and rear sides of the target in the case of free boundary conditions when the energy conservation law was applied for controlling the accuracy of the calculations. A spatial resolution of $\unit[1]{\mu m}$ in $r$ and \unit[1]{nm} in $z$ direction is used. To test the accuracy of the model the energy conservation law was utilized for the case of fixed boundary conditions at the front, rear and lateral sides. 

%% file: comparison.tex
\section{Comparison of Simulation with Experiment and Discussion}
\begin{figure}[h]
 	\centering
 	\includegraphics[width=0.49\textwidth]{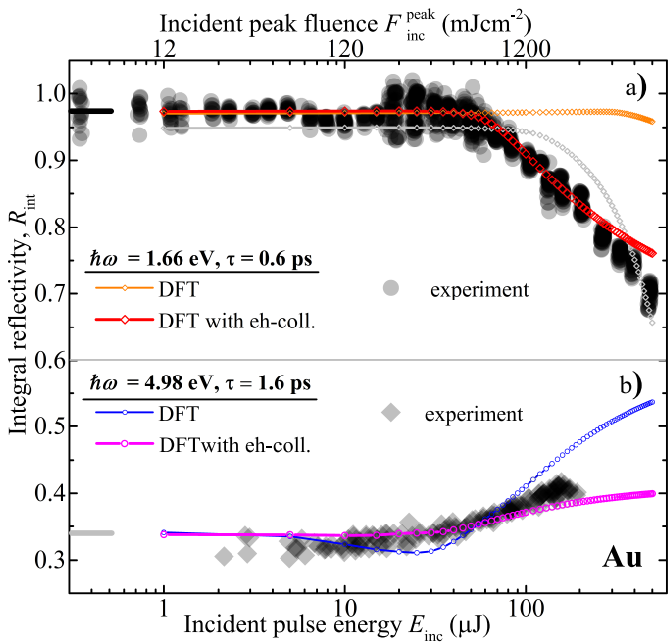}
 	\caption{Single-shot self-reflectivity of gold measured at different incident pulse energies for laser pulses with $\unit[\hbar \omega=1.66]{eV}$ (IR) and $\tau=\unit[0.6]{ps}$ (diamonds in red and orange) in a), $\unit[\hbar \omega=4.98]{eV}$ (UV) and $\tau=\unit[1.6]{ps}$ (circles in blue and pink) in b) respectively, focused on a spot with $\unit[b=85]{\mu m}$ compared to simulations obtained by using temperature-dependent DFT calculations, combined with the effect of eh-collisions. 
	 The literature values of Johnson and Christy for low pulse energies are $R^{Au}_{1.66eV}=0.974$ (black bar) and $R^{Au}_{4.98eV}=0.340$ (gray bar) \cite{johnson1972optical}.}
 	\label{fig:reflectivity}
 \end{figure}

The simulated integral values of reflectivity described by Eq.\,(\ref{eq:reflect}) are directly compared to the experimental results of the self-reflectivity by using the TTM and implementing both of the reflectivity maps $R(T_e, \hbar \omega)$ shown in Fig. \ref{fig:R_map} a) and b). The results are plotted In Fig.~\ref{fig:reflectivity} versus the incident pulse energies resulting from the spatial and temporal integration showing atop the corresponding incident peak fluence. A change in reflectivity is observed in both experiment and simulation above $\unit[F_\mathrm{inc}^\mathrm{peak}=600]{mJ cm^{-2}}$ in the IR for the \unit[0.6]{ps} laser pulses and above $\unit[F_\mathrm{inc}^\mathrm{peak}=120]{mJ cm^{-2}}$ for the \unit[1.6]{ps} pulses in the UV when looking at the simulation where DFT with eh-collisions are implemented, shown in (red diamonds) and (magenta circles), respectively. The difference of the onset of the reflectivity change in the IR and UV can be explained by the initial reflectivity difference at equilibrium conditions and therefore the difference of the absorbed amount of energy and thus the reached electronic temperatures. Even though the rise of R in UV appears at much higher $T_e$ than the drop in R in the probed IR range. When looking at the simulation results where only the ab-inito results of the modified WIEN2k code are used shown in (orange diamonds) and (blue circles), for IR and UV respectively, it appears that the shift of the chemical potential and the broadening of the Fermi-Dirac distribution alone is not able to describe precisely the change in reflectivity observed in experiment. To describe the strong decrease of reflectivity in Fig.~\ref{fig:reflectivity} in the IR and the only small increase of R in the UV a temperature-dependent eh-collision obtained from the calculated DOS for the intra-band excitations needs to be included, resulting in an remarkable agreement of experiment and simulation at both probed wavelength even though one has to mention that therfore a parameter $A_{eh}$ to the ab-initio calculations needs to be included. 

The electronic temperature thus seems to be the key parameter describing the optical response around a fluence relevant for metal surface structuring. In the IR the rise of $T_e$ explains an increases of the rate of free electrons colliding with bound states of the d-band, where the depopulation of states by a broadened Fermi-Dirac distribution increases the probability of free electrons to collide with. Also the effect of the broadening of the excitation edge of the d-band itself plays a role as shown in the DFT results in In Fig. \ref{fig:R_map} a). In the UV an increase of reflectivity is observed in experiment and the DFT simulations described in section~\ref{sec:DFT} and suggest that in this case mainly the extent of the broadening of the Fermi-Dirac distribution explains the change in reflectivity. The explanation is that a de-population of the d-band which creates more occupied states in the s/p-band and leads to an effective shift of the now broad chemical potential $\mu(T_e)$ to higher energies and thus also increases the relative depth of the d-band. The UV photons thus have an increased probability to excite from a bound state since excitations from bound states to unoccupied states even at an excitation energy of $\unit[5]{eV}$ is possible.

An increase of the electronic temperature up to $\unit[T_e=4]{kK}$ will already change the reflectivity around the absorption edge and is referred to as thermo-reflectance \cite{hohlfeld2000electron,sun1994femtosecond,schoenlein1987femtosecond}. In literature the effect is normally described by a simplified picture of a smearing of the excitation from the Fermi level $E_F$ (chemical potential $\mu$) to a sharp d-band \cite{hohlfeld2000electron}. At these elevated $T_e$ no material changes after equilibrating with the lattice will appear. However, when the laser induced energy is sufficient to ablate material after its transfer to the lattice electronic temperatures up to $\unit[80]{kK}$ can be reached during transient self-reflectance. This state is often referred to as warm dense matter \cite{ping2010warm,chen2013evolution}. Under these conditions the density of states (DOS) itself changes, the occupation around the chemical potential $\mu$ smears out spanning a few $\unit[]{eV}$ on the photon energy axis and a shift of $\mu$ is observed \cite{lin2008electron,holst2014ab}. At these conditions a dynamic decrease in reflectivity during the pulses interaction with the surface can even produce a self strengthening effect altering the onset of the ablation threshold of a pulse and the introduced amount of energy. Parameters crucial for a precise simulation of the nano-structuring of metal surfaces \cite{ivanov2015experimental,wu2014microscopic}.


%% file: conclusion.tex
\section{Conclusion}
The reflectivity change under electron-phonon non-equilibrium conditions is measured, simulated and described for a wide range of photon energies at pulse energies relevant for structuring purposes. A map describing $R(T_e, \hbar \omega)$ is introduced and visualises the effect of a boradeining of the Fermi-Dirac distribution on the typical d-band absorption edge of gold as well as the effect of introducing an additional $T_e$ dependet damping factor on the IR reflectivity of gold. At two probe wavelength in the IR and UV a good agreement between experiment and simulation was shown. Our model is also in agreement with experimental thermo-reflectance data around the absorption edge described by Hohlfeld et al. and others \cite{hohlfeld2000electron,sun1994femtosecond,schoenlein1987femtosecond}, and gives a more detailed picture of the involved processes than described before. Agreeing also with experiments at warm dense matter conditions, a decrease in reflectivity in the IR described by Fourment et al. and others \cite{fourment2014exp, zhang2015modeling}. In the UV the increase in reflectivity described by Fedosejevs et al. \cite{fedosejevs1990absorption} also agrees qualitatively with our model. The effect is distinguished from effects related to the formation of a plasma mirror. The relevant phenomena assumed here appearing at elevated electronic temperatures are a smearing of the excitation into the d-band combined with a eh-collision rate increase of free electrons with bound d-band holes being responsible for the dynamic change in reflectivity around the ablation regime, especially in the IR. The shown approach represents a powerful tool, allowing the description of the most general case of laser self-reflectivity and its precise absorbed energy at a certain time and location for elevated electronic temperatures.